# Temporal and Thermal Stability of Al$_2$O$_3$-passivated Phosphorene MOSFETs

Xi Luo, *Student Member, IEEE,* Yaghoob Rahbarihagh, James C. M. Hwang, *Fellow, IEEE,* Han Liu, Yuchen Du, and Peide D. Ye, *Fellow, IEEE*

*Abstract*—This letter evaluates temporal and thermal stability of a state-of-the-art few-layer phosphorene MOSFET with Al$_2$O$_3$ surface passivation and Ti/Au top gate. As fabricated, the phosphorene MOSFET was stable in atmosphere for at least 100 h. With annealing at 200°C in dry nitrogen for 1 h, its drain current increased by an order of magnitude to approximately 100 mA/mm, which could be attributed to the reduction of trapped charge in Al$_2$O$_3$ and/or Schottky barrier at the source and drain contacts. Thereafter, the drain current was stable between −50°C and 150°C up to at least 1000 h. These promising results suggest that environmental protection of phosphorene should not be a major concern, and passivation of phosphorene should focus on its effect on electronic control and transport as in conventional silicon MOSFETs. With cutoff frequencies approaching the gigahertz range, the present phosphorene MOSFET, although far from being optimized, can meet the frequency and stability requirements of most flexible electronics for which phosphorene is intrinsically advantageous due to its corrugated lattice structure.

## I. INTRODUCTION

PHOSPHORENE, a new two-dimensional atomic-layer semiconductor with sizable intrinsic bandgap and carrier mobility is a promising transistor material. To date, most phosphorene MOSFETs were realized [1]-[6] in phosphorene layers exfoliated from black phosphorus and transferred onto SiO$_2$-coated and degenerately doped silicon substrates, which served as gate electrodes on the bottom. The top surface of phosphorene was then metalized with source and drain electrodes in selected areas, leaving the phosphorene channel either exposed or protected by only a thin polymer layer. However, although there are no dangling bonds on the surface, phosphorene is hydrophilic due to an out-of-plane dipole moment [7]. Within hours of exposure to ambient air, the phosphorene surface can become rough [2] and covered by water droplets [8]. Thus, although phosphorene is sufficiently stable for MOSFETs to be fabricated and tested quickly, proper surface passivation is paramount for long-term stability.

Most recently, phosphorene MOSFETs were fabricated [9] with Al$_2$O$_3$ surface passivation and Ti/Au top gate. However, as the process was not optimized, the performance of these MOSFETs was impacted by trapped charge in the Al$_2$O$_3$ passivation and Schottky barriers at the source and drain contacts. This letter further evaluates their effects on temporal and thermal stability. Other than replacing the SiO$_2$-coated Si conducting substrate with an Al$_2$O$_3$-coated GaAs semi-insulating substrate, the present process for Al$_2$O$_3$ surface passivation and Ti/Au top gate is exactly the same as that of [9] except for random process deviations typical of academic research. The GaAs semi-insulating substrate reduces parasitic capacitances and allows RF performance characteristics to be evaluated for the first time.

## II. EXPERIMENT

The fabrication of phosphorene MOSFETs started with few-layer phosphorene approximately 5-nm thick that was exfoliated from bulk black phosphorus then transferred onto a semi-insulating GaAs substrate coated with 50 nm of atomic-layer deposited (ALD) Al$_2$O$_3$. The transferred phosphorene was then metallized with source and drain electrodes in regions defined by electron-beam lithography to be 1-μm apart. The electrodes consisted of sequentially evaproated Ni and Au layers with thicknesses of 20 nm and 60 nm, respectively. Following source and drain metalization, the phosphorene surface was passivated by ALD Al$_2$O$_3$. The ALD was seeded by a 0.8-nm-thick pure Al layer, which was expected to getter oxygen from the phosphorene surface as well as the atmosphere and to become completely oxidized before ALD. The ALD was carried out with trimethylaluminum and water precursors at 250°C until 15 nm of Al$_2$O$_3$ was deposited. Following surface passivation, a gate electrode was defined by electron-beam lithography in the middle of the source-drain spacing with a length of 0.7 μm. The gate width was 3 μm. The gate electrode

Manuscript submitted August 12, 2014; revised September 15, 2014 and October 1, 2014; accepted October 3, 2014. This work was supported in part by the US Department of Defense, Office of Naval Research under Grant N00014-14-1-0653, and the National Science Foundation under Grant ECCS-1449270.

X. Luo, Y. Rahbarihagh, and J. C. M. Hwang are with the Department of Electrical and Computer Engineering, Lehigh University, Bethlehem, PA 18015 USA (e-mail: xil609@lehigh.edu).

H. Liu, Y. Du, and P. D. Ye are with Birck Nanotechnology Center, School of Electrical and Computer Engineering, Purdue University, West Lafayette, IN 47907 USA.

Color versions of one or more of the figures in this letter are available online at http://ieeexplore.ieee.org.

Digital Object Identifier XXX.

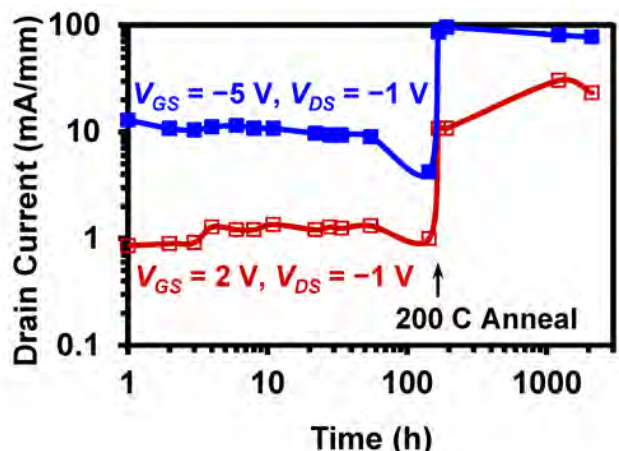

Fig. 1. Measured drain current drifts of a typical phosphorene MOSFET with $Al_2O_3$ surface passivation and Ti/Au top gate. The "on" and "off" currents are indicated by filled and empty symbols, respectively.

consisted of sequentially evaproated Ti and Au layers with thicknesses of 20 nm and 60 nm, respectively. For parasitics de-embedding, a gateless MOSFET with only the gate interconnect and probe pad was fabricated along with the gated MOSFET.

After the phosphorene MOSFET was fabricated, it was characterized in a Cascade Microtech Summit 1000 thermal probe station equipped with an Attoguard MicroChamber for enviromental control. DC characterization was performed by using an Agilent Technologies 4156C precision semiconductor parameter analyzer. RF characterization was performed by using an Agilent Technologies N5230A PNA network analyzer. Room-temperature DC and RF characterization was conducted in room air, whereas annealing and temperature-dependent characterization were conducted in dry nitrogen.

### III. RESULT AND DISCUSSION

Fig. 1 shows the temporal stability of a typical phosphorene MOSFET with $Al_2O_3$ passivation and Ti/Au gate. As fabricated, its "on" and "off" currents were generally stable in room air and temperature for more than 100 h. After 144 h, the MOSFET was annealed at 200°C in dry nitrogen for 1 h, resulting in an order-of-magnitude increase of the "on" current. This increase appeared to be permanent to at least 1000 h in room air and temperature, and could be attributed to the reduction of trapped charge in $Al_2O_3$ and/or Schottky barrier at the source and drain contacts [9]. As shown in Fig. 2, the "on" current measured at a gate-source voltage $V_{GS}$ of −5 V and a drain-source voltage $V_{DS}$ of −1 V can continue to increase with $V_{GS} < -5$ V, whereas the "off" current measured at $V_{GS} = 2$ V and $V_{DS} = -1$ V can continue to decrease with $V_{GS} > 2$ V. Thus, these apparent "on" and "off" currents are convenient stability measures but not true measures of switching capacity.

Fig. 2 and Fig. 3 show the DC and RF characteristics, respectively, measured on the same annealed phosphorene MOSFET as shown in Fig. 1. Additionally, Fig. 2 includes the DC characteristics before annealing which were amplified ten times to facilitate comparison with the characteristics after annealing. It can be seen that the MOSFET exhibits a p-type behavior with a saturated drain current on the order of 100 mA/mm and a peak transconductance on the order of 100 mS/mm. Most of the drain current originates from the source, as the measured gate current is on the order of picoampere. In Fig. 3, it can be seen that the de-embedded forward-current gain cutoff frequency $f_T$ and the maximum frequency of oscillation $f_{MAX}$ approach the gigahertz range. This suggests that, the present phosphorene MOSFET, while far below the theoretically projected performance limit [10], can already meet the frequency and stability requirements of most flexible electronics such as driver circuits for flat-panel displays [11]. (Phosphorene is intrinsically flexible due to its corrugated lattice structure [1].) Based on order-of-magnitude estimates, the present DC and RF characteristics are consistent with that of a MOSFET with a gate length of 1 μm, a gate width of 10 μm, a carrier density of $10^{13}$ cm$^{-2}$, a gate capacitance of 10 pF/mm, a field mobility of 100 cm$^2$/V·s, a drain transconductance of 100 mS/mm, and a contact resistance of 10 Ω·mm. In turn, these properties are consistent with the state of the art of phosphorene MOSFETs [1]-[6] with ample room for improvement.

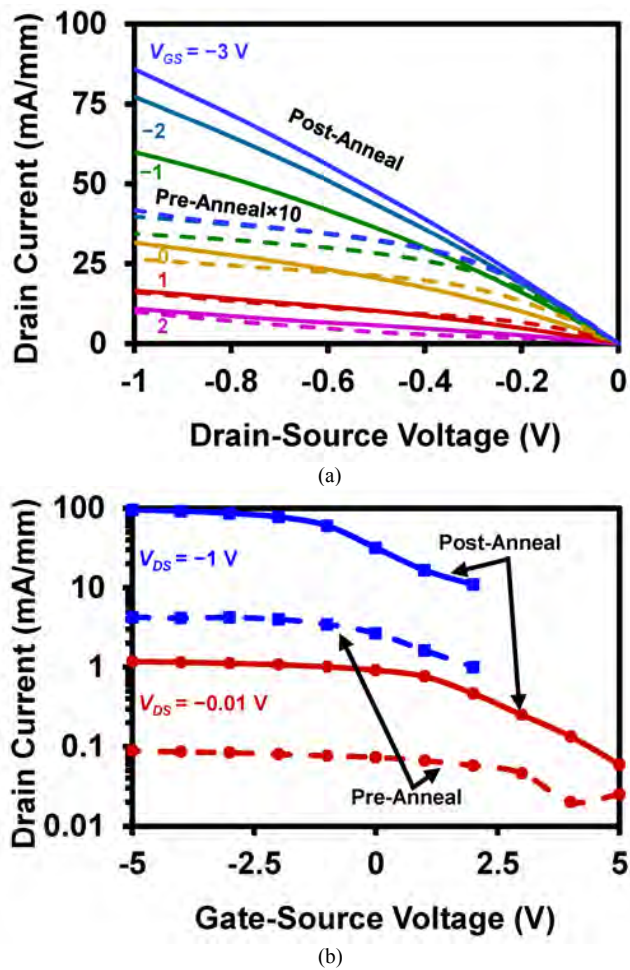

(a)

(b)

Fig. 2. Measured (a) drain and (b) transfer characteristics of the phosphorene MOSFET before (dashed curves) and after (solid curves) annealing at 200 °C in dry nitrogen for 1 h. In (a), pre-anneal currents are amplified 10X so as to appear on the same linear scale as post-anneal currents. $V_{GS} = -3, -2 \ldots 2$ top down. In (b), "linear" and "saturated" trassfer characteristics measued at $V_{DS} = -0.01$ V and −1 V are indicated by round and square symbols, respectively.

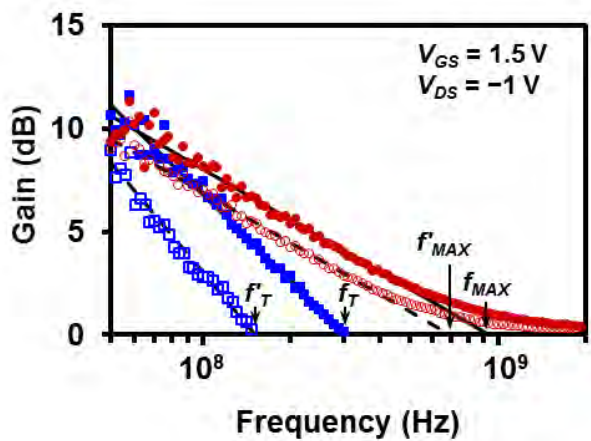

Fig. 3. De-embedded forward current gain cutoff frequency $f_T$ and maximum frequency of oscillation $f_{MAX}$ of the annealed phosphorene MOSFET. As-meassured values are indicated as $f'_T$ and $f'_{MAX}$.

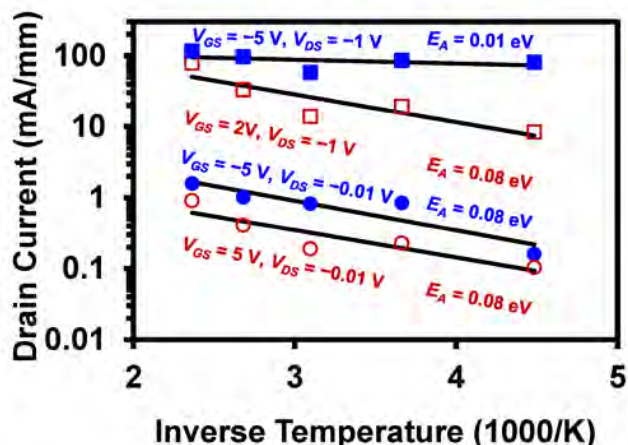

Fig. 4. Measured temperature dependence between −50 °C and 150 °C of both linear and saturated "on" and "off" drain currents of the annealed phosphorene MOSFET.

De-embedding from the as-measured $S$ parameters was necessary, because the parasitic capacitance measured on the gateless MOSFET was approximately 22 fF, which was on the same order as the gate capacitance of the intrinsic MOSFET. The de-embedding followed the "cold FET" procedure [12]. First, the measured $S$ parameters were converted to $Y$ parameters so that the shunt capacitance associated with the parasitic capacitance could be subtracted. Second, the remaining $Y$ parameters were used to calculate the maximum available gain $MAG$. Third, the remaining $Y$ parameters were converted to $h$ parameters to arrive at the short-circuit forward-current gain $h_{21}$. Finally, $h_{21}$ and $MAG$ were extrapolated to unity for $f_T$ and $f_{MAX}$, respectively, as shown in Fig. 3.

Fig. 4 shows the thermal stability between −50°C and 150°C of the same annealed phosphorene MOSFET. It can be seen that, both the "on" and "off" currents exhibit an Arrhenius dependence with an activation energy $E_A < 0.1$ eV. In particular, the saturated "on" current measured with $V_{GS}=-5$ V and $V_{DS}=-1$ V has a very weak temperature dependence of $E_A = 0.01$ eV. This is not characteristic of bulk semiconductors with a narrow bandgap, considering the bandgap of 5-nm-thick phosphorene is approximately 0.3 eV [1]. This may be due to the delicate balance of competing mechanisms such as the reduction of the Schottky barrier at the source and drain contacts [9] and carrier mobility in the channel [13]. Such complicated mechanisms can be separated only after more thorough study.

## IV. CONCLUSION

Compared to previously reported phosphorene MOSFETs without surface passivation, the present $Al_2O_3$-passivated phosphorene MOSFET appears to be orders-of-magnitude more stable both temporally and thermally. This should help alleviate concerns over the oxidation tendency of phosphorene. However, the present $Al_2O_3$ passivation contains trapped charge that can significantly impact electronic control and transport. This will require more detailed passivation study and optimization as in the decades-long development of Si MOSFETs. Eventually, as another elemental semiconductor with simpler chemistry than that of compound semiconductors, phosphorene MOSFETs may be passivated as well as Si MOSFETs are now.